\documentclass[prb,aps,twocolumn,floatfix]{revtex4}

\usepackage[pdftex]{graphicx}
\usepackage{amsmath,amsfonts,amsbsy,mathrsfs,amssymb}

\begin{document}
\title{Group delay in THz spectroscopy with ultra-wideband log-spiral antennae}

\author{M. Langenbach,$^{1}$ A. Roggenbuck,$^{2}$ I.~C{\'a}mara Mayorga,$^{3}$ A. Deninger,$^{2}$
K. Thirunavukkuarasu,$^{1,4}$ J. Hemberger,$^{1}$ and M.~Gr{\"u}ninger$^{1}$}
\affiliation{$^1$ II. Physikalisches Institut, Universit\"{a}t zu K\"{o}ln,
Z\"{u}lpicher Str.\ 77, D-50937 K\"{o}ln, Germany\\
$^2$ TOPTICA Photonics AG, Lochhamer Schlag 19, D-82166 Gr\"{a}felfing, Germany\\
$^3$ Max-Planck-Institute for Radio Astronomy, Auf dem H\"{u}gel 69, D-53121 Bonn, Germany\\
$^4$ National High Magnetic Field Laboratory, Tallahassee, Florida 32310, USA}

\begin{abstract}
We report on the group delay observed in continuous-wave terahertz spectroscopy based on photomixing
with phase-sensitive homodyne detection.
We discuss the different contributions of the experimental setup to the phase difference $\Delta \varphi(\nu)$
between transmitter arm and receiver arm.
A simple model based on three contributions yields a quantitative description of the
overall behavior of $\Delta \varphi(\nu)$.
Firstly, the optical path-length difference gives rise to a term linear in frequency $\nu$.
Secondly, the ultra-wideband log-spiral antennae effectively radiate and receive in a frequency-dependent
active region, which in the most simple model is an annular area with a circumference equal to the wavelength.
The corresponding term changes by roughly $6\pi$ between 100\,GHz and 1\,THz.
The third contribution stems from the photomixer impedance.
In contrast, the derivative $\partial \Delta \varphi/\partial \nu$ is dominated by
the contribution of periodic modulations of $\Delta \varphi(\nu)$ caused by standing waves,
e.g., in the photomixers' Si lenses.
Furthermore, we discuss the Fourier-transformed spectra, which are equivalent to the waveform
in a time-domain experiment. In the time domain, the group delay introduced by the log-spiral antennae
gives rise to strongly chirped signals, in which low frequencies are delayed.
Correcting for the contributions of antennae and photomixers yields sharp peaks or ``pulses'' and
thus facilitates a time-domain-like analysis of our continuous-wave data.
\end{abstract}

\date{June 25, 2014}
\maketitle

\section{Introduction}

Continuous-wave (cw) terahertz spectroscopy based on photomixing is able to cover a very broad frequency range
from about 0.1\,THz up to 5\,THz.\cite{McIntosh95}
For broadband spectroscopy, it is desirable that the photomixers in combination with the antennae
provide a rather smooth spectrum without pronounced resonances,
i.e., a nearly frequency-independent radiation pattern and a nearly frequency-independent radiation resistance.
This can be achieved by using a self-complementary antenna such as the log-spiral (or equiangular spiral) antenna,\cite{Wiesbeck09}
which offers a large bandwidth in combination with a high terahertz efficiency and an excellent beam pattern.\cite{Brown95,McIntosh96,Nguyen12}
However, the log-spiral antenna effectively radiates and receives terahertz waves from the frequency-dependent
annular ``active region'' \cite{Kaiser60,McFadden07,McFadden09} with a circumference roughly equal to the wavelength $\lambda$.
Therefore this antenna shows a pronounced frequency dependence of the group delay
\begin{equation}
t_{\rm gr}(\nu) = \frac{1}{2\pi}  \frac{\partial \varphi_{\rm an}}{\partial \nu}
\label{eq:group}
\end{equation}
where $\varphi_{\rm an}(\nu)$ denotes the phase of the wave emitted at frequency $\nu$.
The group delay of the antenna corresponds to the traveling time of the photocurrent
from the inner feed to the active region. This delay may vary strongly over the useable frequency range
of the antenna. In our case, it varies by more than a factor of 10 between 0.1\,THz and 1\,THz.
Accordingly, log-spiral antennae are not well suited for experiments in the time domain,
as a log-spiral antenna fed with a narrow pulse emits a strongly chirped signal.

In frequency-domain terahertz spectroscopy based on homodyne detection,
we measure the phase difference $\Delta \varphi(\nu)$ between transmitter arm and receiver arm, see below.
The phase shift $\phi_{\rm sam}$ introduced by a given sample is determined by comparison with the data
measured in a reference run without sample,
\begin{equation}
\phi_{\rm sam}(\nu) = \Delta \varphi_{\rm with}(\nu) - \Delta \varphi_{\rm w/o}(\nu) \, .
\label{eq:with}
\end{equation}
In an ideal case, the group delay introduced by the antennae is identical in both terms on the right hand side,
hence it does not contribute to $\phi_{\rm sam}(\nu)$.
Nevertheless, it is instructive to quantitatively understand the phase difference
$\Delta \varphi_{\rm w/o}(\nu)$ measured in the reference run, e.g., for a discussion of the
uncertainty of the phase caused by a drift of the frequency, cf.\ Sec.\ \ref{sec:gr}.
Moreover, a quantitative description of the reference phase allows for a correction of the
frequency dependence of the group delay
and thus facilitates a time-domain-like analysis of the cw data.
To the best of our knowledge, the group delay of photomixers with log-spiral antennae working in the tera\-hertz range
has not been reported thus far.
Here, we systematically discuss all contributions to the frequency dependence of the phase difference $\Delta \varphi(\nu)$.
We employ a simple model and obtain a quantitative description of the overall behavior of $\Delta \varphi(\nu)$.

\section{Experimental setup}

\begin{figure}[tb]
\centering
\includegraphics[width=.95\columnwidth]{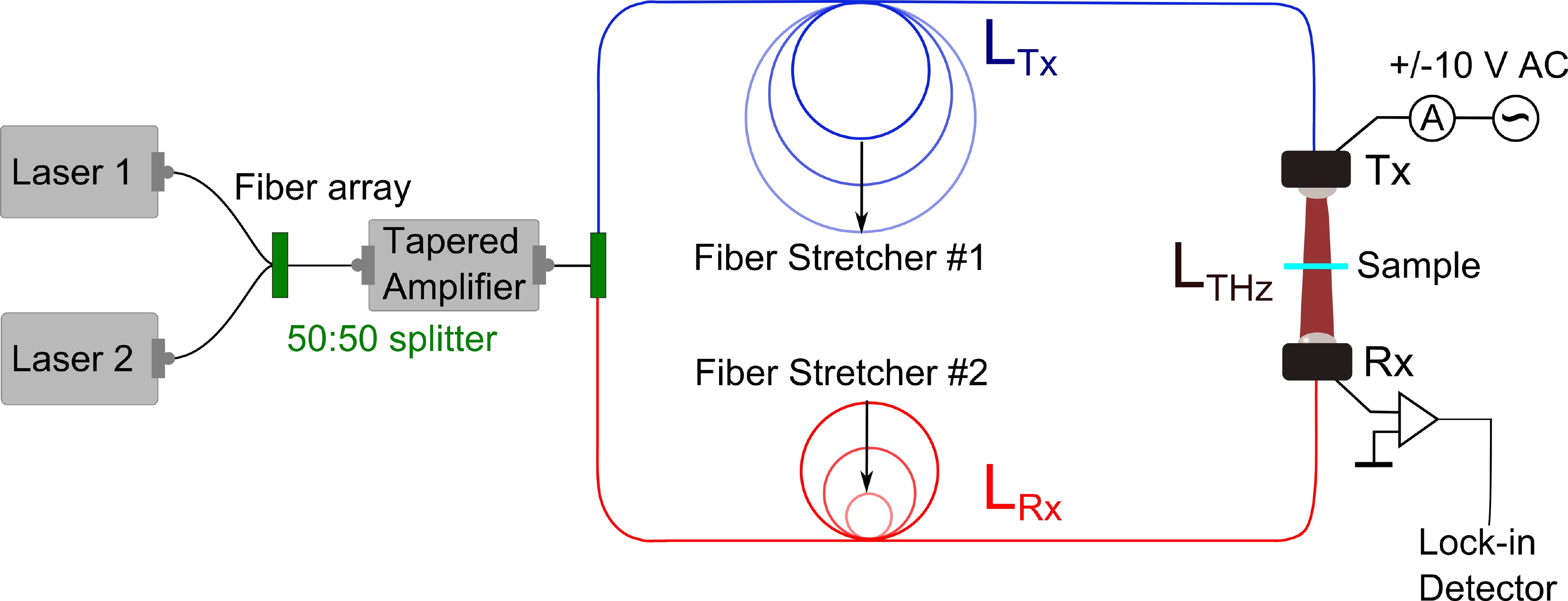}
\caption{Sketch of the setup. The symbols Tx and Rx refer to transmitter and receiver, respectively.}
\label{fig:Setup}
\end{figure}

A sketch of our experimental setup is given in Fig.\ \ref{fig:Setup}, for details we refer to
Refs.\ [\onlinecite{Deninger08,Roggenbuck10,Roggenbuck12,Roggenbuck13}].
Continuous-wave terahertz radiation with frequency $\nu$\,=\,$|\nu_2-\nu_1|$ is generated and coherently detected
by illuminating two photomixers, transmitter and receiver, with the optical beat of two near-infrared lasers
with frequencies $\nu_1$ and $\nu_2$.
We use two laser diodes with slightly different wavelengths centered at about 780\,nm,
offering a maximum beat frequency of about 1.8\,THz with a line width of about 5\,MHz.
The laser light is guided in a fiber array with two fiber-optical 50:50 splitters. The first splitter is used
to superimpose the two laser beams,
which are subsequently amplified in a tapered semiconductor amplifier.
The second fiber-optical splitter separates the transmitter arm and the receiver arm.

In order to obtain information on both amplitude and phase, we employ fast phase modulation via
two fiber stretchers\cite{Roggenbuck12}
in the optical path before the photomixers, i.e., where both laser frequencies are superimposed.
The two stretchers operate with opposite signs, thus changing the optical path-length difference
\begin{equation}
\Delta L  =  L_{\rm Tx} + L_{\rm THz} - L_{\rm Rx}
\label{eq:DL}
\end{equation}
between the transmitter arm including the THz path with the total optical path length $L_{\rm Tx} + L_{\rm THz}$
on the one side, and the receiver arm with the optical path length $L_{\rm R_x}$ on the other side.

\begin{figure}[tb]
\centering
\includegraphics[width=.5\columnwidth]{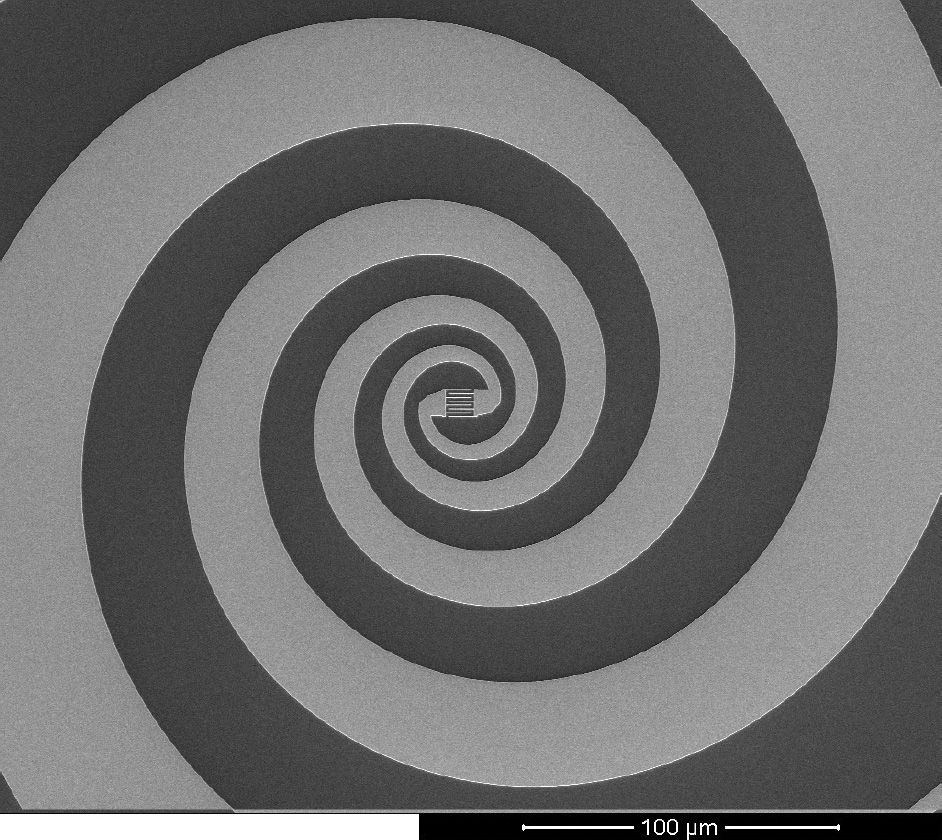}
\caption{Log-spiral antenna and interdigitated finger structure.}
\label{fig:antenna}
\end{figure}

The photomixers are based on ion-implanted GaAs and have been described in Ref.\ [\onlinecite{Mayorga07}].
The photomixing area with dimensions of $9\times 9$\,$\mu$m$^2$ consists of an interdigitated
metal-semiconductor-metal structure with eight fingers, see Fig.~\ref{fig:antenna}.
The metallization consists of a 10/200\,nm thick Ti/Au layer.
The patterned antennae are self-complementary log-periodic spirals with three turns.
The spiral radius $r(\alpha)$ as a function of the angle $\alpha$ is described by
\begin{equation}
r(\alpha) = r_{\rm min}\, e^{a\alpha}
\label{eq:spiral}
\end{equation}
with the minimum radius $r_{\rm min} \! \approx \! 10$\,$\mu$m and growth rate $a$\,=\,0.2.
With three turns, the maximum radius amounts to $r_{\rm max} \! \approx$\,0.43\,mm.
The outer spiral antenna arms are bonded in order to bias the photomixer structure
in the case of the transmitter, or to measure the DC photocurrent of the receiver.

Due to the large dielectric constant of $\varepsilon$(GaAs)\,=\,12.8, the antenna radiates mainly into the substrate.
For an efficient coupling to free space, each photomixer is mounted on a hyper-hemispherical lens made of high-resistivity Si.
Terahertz radiation is emitted with a full opening angle of only $10^\circ$ at 100\,GHz, $4^\circ$ at 350\,GHz, and $\leq\!
2^\circ$ between 600\,GHz and 1.2\,THz.\cite{Roggenbuck12}
For a short distance of $L_{\rm THz} \! \lesssim \! 30$\,cm between the two photomixers, this allows us
to employ a face-to-face configuration without any further focussing optics.

\section{Results}

Based on homodyne detection, the photocurrent $I_{\rm ph}$ in the receiver is given by\cite{Verghese98}
\begin{equation}
    I_{\rm ph}\propto E_{\rm THz}\cos(\Delta\varphi) \, ,
\label{eq:Iph}
\end{equation}
where $E_{\rm THz}$ denotes the amplitude of the incident terahertz electric field and $\Delta \varphi$
the phase difference between the optical signal and the terahertz signal at the receiver.
Experimentally, $\Delta \varphi$ is determined only up to an offset $m \cdot 2\pi$,
where $m$ is an integer number. However, to reveal the optical properties of a given
sample we have to consider $m_{\rm with} - m_{\rm w/o}$, i.e., the difference between sample and
reference run (cf.\ Eq.\ \ref{eq:with}). By measuring over a broad frequency range and comparison with
the model derived below, the ambiguity of $m$ can be resolved.

\begin{figure}[tb]
\centering
\includegraphics[width=.9\columnwidth]{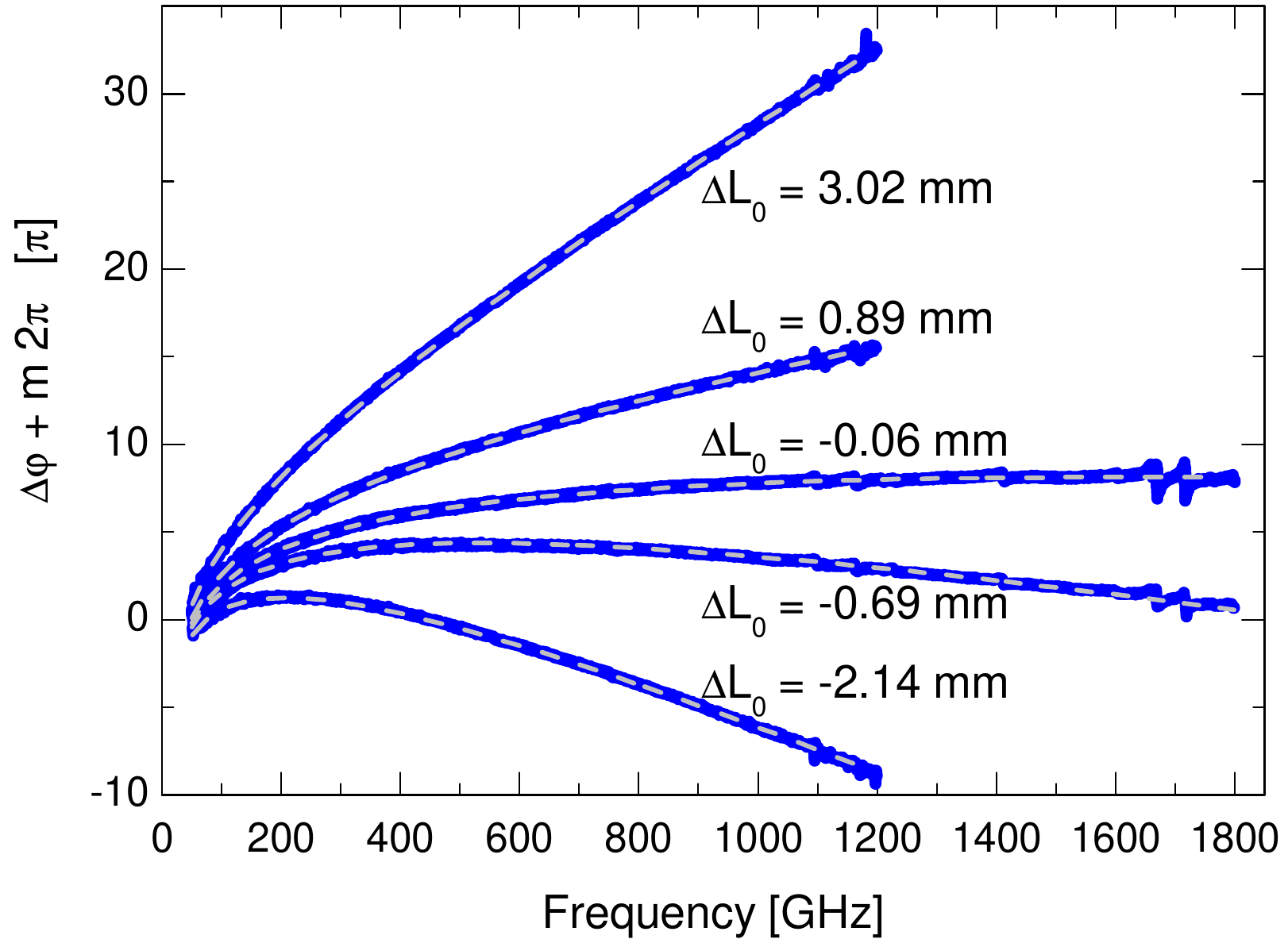}
\caption{Blue: Representative data sets of $\Delta \varphi(\nu)$ measured in a face-to-face configuration
of the two photomixers. Different values of the optical path-length difference
$\Delta L$ were obtained by changing $L_{\rm THz} \! \approx \! 22$\,cm.
The frequency step width equals 100\,MHz for $\Delta L \! \approx \! -0.06$\,mm and 1\,GHz for all other data sets.
Grey: calculated values of $\Delta \varphi_{\rm mod}(\nu)$ according to Eq.\ \ref{eq:model}
with the fit parameters $\Delta \varphi_{\rm an}(\nu_{\rm max})$\,=\,0.27\,$\pi$
and $\Delta L_0$ as given in the plot.
The comparison between model and measured data yields $m\cdot 2\pi$\,=\,8$\pi$.
}
\label{fig:phase}
\end{figure}

Representative data of $\Delta \varphi(\nu)$ measured for different values of $\Delta L$ are depicted in Fig.\ \ref{fig:phase}.
These data sets were obtained in reference runs without any sample.
The dominant behavior at high frequencies is linear in frequency.
At low frequencies, we observe strong deviations from linearity, which is most obvious for small values of $\Delta L$.

The frequency dependence of $\Delta \varphi(\nu)$ is related to the group delay difference $\Delta t_{\rm gr}$ by
\begin{equation}
\Delta t_{\rm gr} = \frac{1}{2\pi} \frac{\partial \Delta \varphi(\nu)}{\partial \nu} \, ,
\label{eq:tgr}
\end{equation}
where $\Delta t_{\rm gr}(\nu)$ describes the difference in traveling time between
transmitter arm and receiver arm for a wave packet centered at $\nu$.
Let us first consider the most simple case without dispersion, i.e., a frequency-independent propagation velocity $c$
in combination with a frequency-independent optical path-length difference $\Delta L_0$
without any further group delays.
Then, $\Delta t_{\rm gr}(\nu)$\,=\,$\Delta L_0 / c$ is independent of frequency, giving rise to
a linear behavior $\Delta \varphi \! \propto \! \nu$.
In the following, we systematically address all contributions to $\Delta \varphi(\nu)$:
(1) the fibers,
(2) the photomixers including the antennae and the hyper-hemispherical Si lenses, and
(3) the terahertz path $L_{\rm THz}$ including, e.g., air with water vapor
or standing waves between, e.g., the photomixers.

\textit{(1) Fibers:}
The refractive index $n_{\rm f}$ of a fiber of length $L_{\rm f}$ depends on the frequencies
of the two near-infrared lasers.
A terahertz frequency $\nu$\,=\,$\nu_2 - \nu_1$ is selected by scanning the two laser frequencies symmetrically
around the center frequency $\nu_0$\,=\,$(\nu_1 + \nu_2)/2$,
i.e., $\nu_{2,1}$\,=\,$\nu_0 \pm \nu/2$. The phase of the optical beat is given by
\begin{equation}
 \varphi = L_{\rm f} \cdot \left[ n_{\rm f}(\nu_2)\cdot \nu_2 - n_{\rm f}(\nu_1)\cdot \nu_1 \right] \cdot \frac{2\pi}{c} \, .
\label{eq:beatphase}
\end{equation}
We assume that the dispersion is linear around 780\,nm, which indeed is the case for the fiber material SiO$_2$
with $\partial n/\partial \nu$\,=\,4$\cdot 10^{-5}/$THz.\cite{Malitson65}
We expand around the center frequency $\nu_0$,
\begin{equation}
 n_{\rm f}(\nu_{2}) = n_{\rm f}(\nu_0) + \left. \frac{\partial n_{\rm f}}{\partial \nu}\right|_{\nu_0} \cdot (\nu_2 - \nu_0)
= n_{\rm f}(\nu_0) + \left. \frac{\partial n_{\rm f}}{\partial \nu}\right|_{\nu_0} \cdot \frac{\nu}{2}  \, ,
\label{eq:n12}
\end{equation}
\begin{equation}
 n_{\rm f}(\nu_{1}) = n_{\rm f}(\nu_0) + \left. \frac{\partial n_{\rm f}}{\partial \nu}\right|_{\nu_0} \cdot (\nu_1 - \nu_0)
= n_{\rm f}(\nu_0) - \left. \frac{\partial n_{\rm f}}{\partial \nu}\right|_{\nu_0} \cdot \frac{\nu}{2}  \, ,
\end{equation}
and find
\begin{equation}
 \varphi = L_{\rm f} \cdot \left(n_{\rm f}(\nu_0)
 + \left. \frac{\partial n_{\rm f}}{\partial \nu}\right|_{\nu_0} \cdot \nu_0  \right) \cdot \frac{2\pi\nu}{c}
= \, L_{\rm f} \cdot n_{\rm f,eff} \cdot \frac{2\pi\nu}{c} \, .
\end{equation}
The term in parentheses is independent of frequency. Due to the fact that the two lasers are scanned in opposite directions,
the linear frequency dependence of $n_{\rm f}$ simply gives rise to a slight increase $n_{\rm f,eff}/n_{\rm f} \! \approx \! 1.01$
of the \textit{frequency-independent} effective fiber length $n_{\rm f,eff} \cdot L_{\rm f}$
relevant for the optical beat. The contribution of the fibers thus reads
\begin{equation}
 \Delta \varphi_{\rm f} = \left( L_{\rm Tx} - L_{\rm Rx} \right) \cdot \frac{2\pi\nu}{c}  \, ,
\label{eq:beatphase_2laser}
\end{equation}
where $L_{\rm Rx}$ and $L_{\rm Tx}$ include the effective frequency-independent refractive index $n_{\rm f,eff}$ of the fibers.

\textit{(2) Photomixers and antennae:}
We consider the following contributions:
(i) coupling the optical beat into the photoactive area,
(ii) the photoconductance,
(iii) the photomixer impedance,
(iv) the antenna, and
(v) coupling to free space via a hyper-hemispherical Si lens.

(i) We utilize two identical photomixers.
The effect of coupling the optical beat into the photoactive area is thus identical
in the receiver arm and the transmitter arm. Hence it does not contribute to $\Delta \varphi$.

(ii) The same applies to the photoconductance $G$, which depends on the terahertz frequency due to the finite
carrier lifetime $\tau \! \approx \! 0.5$\,ps.\cite{Mayorga07} This gives rise to a phase shift of
$\varphi_{G}$\,=\,$\arctan(2\pi \nu \tau)$ in \textit{both} receiver and transmitter.\cite{GregoryIEEE05,Tani05}
In the transmitter, the terahertz wave is delayed with respect to the optical beat, which effectively
increases $L_{\rm Tx}$. However, the phase shift in the receiver effectively increases $L_{\rm Rx}$ by the same amount.
Therefore, the photoconductance does not contribute to $\Delta \varphi$.

(iii) The total impedance of photomixer and antenna effectively is described by
a characteristic time constant $\tau_{RC}$\,=\,$R_{\rm A} C$,
where $R_{\rm A}$ $\approx$ 73\,${\rm \Omega}$
denotes the nearly frequency-independent antenna resistivity of the
log-spiral antenna on a GaAs substrate,\cite{Nguyen12,Mayorga07} and $C \! \approx \! 1.5$\,fF is the capacitance
of the interdigitated electrode structure.\cite{Mayorga07}
The time constant $\tau_{RC} \! \approx$\,0.1\,ps
gives rise to a phase shift of $\varphi_{RC} = \arctan(2\pi \nu \tau_{RC})$,
again in both receiver and transmitter. However, this phase shift has to
be attributed to the transmitter arm in both
photomixers. Hence both terms add up and yield a contribution of
\begin{equation}
  \Delta \varphi_{RC} = 2 \arctan(2\pi \nu \tau_{RC}) \, .
\label{eq:atanphi}
\end{equation}

(iv) It is well known that ultra-wideband log-spiral antennae exhibit a strong dispersion and
thus distort short pulses.\cite{Wiesbeck09,McFadden07,McFadden09,McFaddenPhD,Hertel03}
The antenna effectively radiates and receives terahertz waves in an annular ``active region''
with radius $r_{\rm ar}$, the size of which depends on frequency.
Physically, the antenna predominantly radiates where the contributions from the two neighboring
spiral arms interfere constructively,\cite{Kaiser60,McFaddenPhD} see Sec.\ \ref{sec:ar}.
In the limit of a vanishing or very small spiral growth rate $a$ (cf.\ Eq.~\ref{eq:spiral}),
a center-fed log-spiral antenna radiates where the circumference equals the wavelength, $2\pi r_{\rm ar}$\,=\,$\lambda$.
Comparing two waves at high and low frequencies, the low-frequency wave is delayed because it has to travel a longer path
$l(\nu)$ in the antenna.\cite{Hertel03} In time, the signal has to travel for
\begin{equation}
t_{\rm gr,an}(\nu) = l(\nu) \cdot n_{\rm eff} /c
\label{eq:time}
\end{equation}
where $t_{\rm gr,an}$ is the group delay of the antenna
and $n_{\rm eff}$ denotes the effective refractive index.
With $\varepsilon$(GaAs)\,=\,12.8, we use $n_{\rm eff}$\,=\,$\sqrt{(12.8+1)/2}$\,$\approx$\,2.6
for the guided mode at the GaAs-air interface.
The path length in the spiral equals
\begin{eqnarray}
\nonumber  l\left(\alpha(\nu)\right) &=& \int_0^{\alpha} \sqrt{r(\alpha^\prime)^2 + \left( \frac{dr}{d\alpha^\prime}\right)^2}\, d\alpha^\prime  \\
   &=& \frac{\sqrt{1+a^2}}{a}\,\, \left(r_{\rm ar}(\nu) - r_{\rm min}\right) \, .
\label{eq:lengthspiral}
\end{eqnarray}
The inner and outer truncations of the spiral define a minimum and maximum wavelength, respectively.
With the radius $r_{\rm ar}$\,=\,$\lambda/2\pi$ of the active region
and $r_{\rm min}$\,=\,10\,$\mu$m we find
$\nu_{\rm max}$\,=\,$c / (2\pi r_{\rm min}n_{\rm eff})$\,$\approx$\,1.8\,THz
as well as $\nu_{\rm min} \! \approx \! 40$\,GHz and
\begin{equation}
l(\nu)  = \frac{c}{ 2 \pi n_{\rm eff}} \frac{\sqrt{1+a^2}}{a} \cdot \left( \frac{1}{\nu} - \frac{1}{\nu_{\rm max}} \right)  \, .
\label{eq:lnu}
\end{equation}
With Eq.\ \ref{eq:time} we integrate Eq.\ \ref{eq:group} from $\nu_{\rm max}$ to $\nu$,
\begin{eqnarray}
 \nonumber \int_{\nu_{\rm max}}^\nu t_{\rm gr} \, d\nu
  &=& \int_{\nu_{\rm max}}^\nu \frac{\sqrt{1+a^2}}{2\pi a} \cdot \left( \frac{1}{\nu} - \frac{1}{\nu_{\rm max}} \right) \, d\nu \\
   &=& \frac{1}{2\pi} [\varphi_{\rm an}(\nu) - \varphi_{\rm an}(\nu_{\rm max})]   \, .
\label{eq:itegral}
\end{eqnarray}
Moreover, we assume that the delay is identical upon emission and detection.
In both cases, the delay effectively prolongs the transmitter arm,
thus we have to add up the two contributions. This finally yields the contribution of the
antenna characteristics to the phase difference $\Delta \varphi(\nu)$ between the two arms,
\begin{equation}
\Delta \varphi_{\rm an}(\nu) = \frac{2\sqrt{1+a^2}}{a}
[ \ln(\frac{\nu}{\nu_{\rm max}}) - \frac{\nu}{\nu_{\rm max}} +1] + \Delta \varphi_{\rm an}(\nu_{\rm max}) \, ,
\label{eq:antenna}
\end{equation}
where the offset $\Delta \varphi_{\rm an}(\nu_{\rm max})$ is treated as a fit parameter.

(v) Standing waves within the hyper-hemispherical Si lenses give rise to a periodic modulation
of $\Delta \varphi(\nu)$.\cite{Boriskin04,GoebelPhD} This effect can be neglected for the discussion of
the overall behavior of $\Delta \varphi(\nu)$. However, these standing waves are important if one
considers the derivative $\partial \Delta \varphi/\partial \nu$, see Sec.\ \ref{sec:gr}.

\begin{figure}[tb]
\centering
\includegraphics[width=.8\columnwidth]{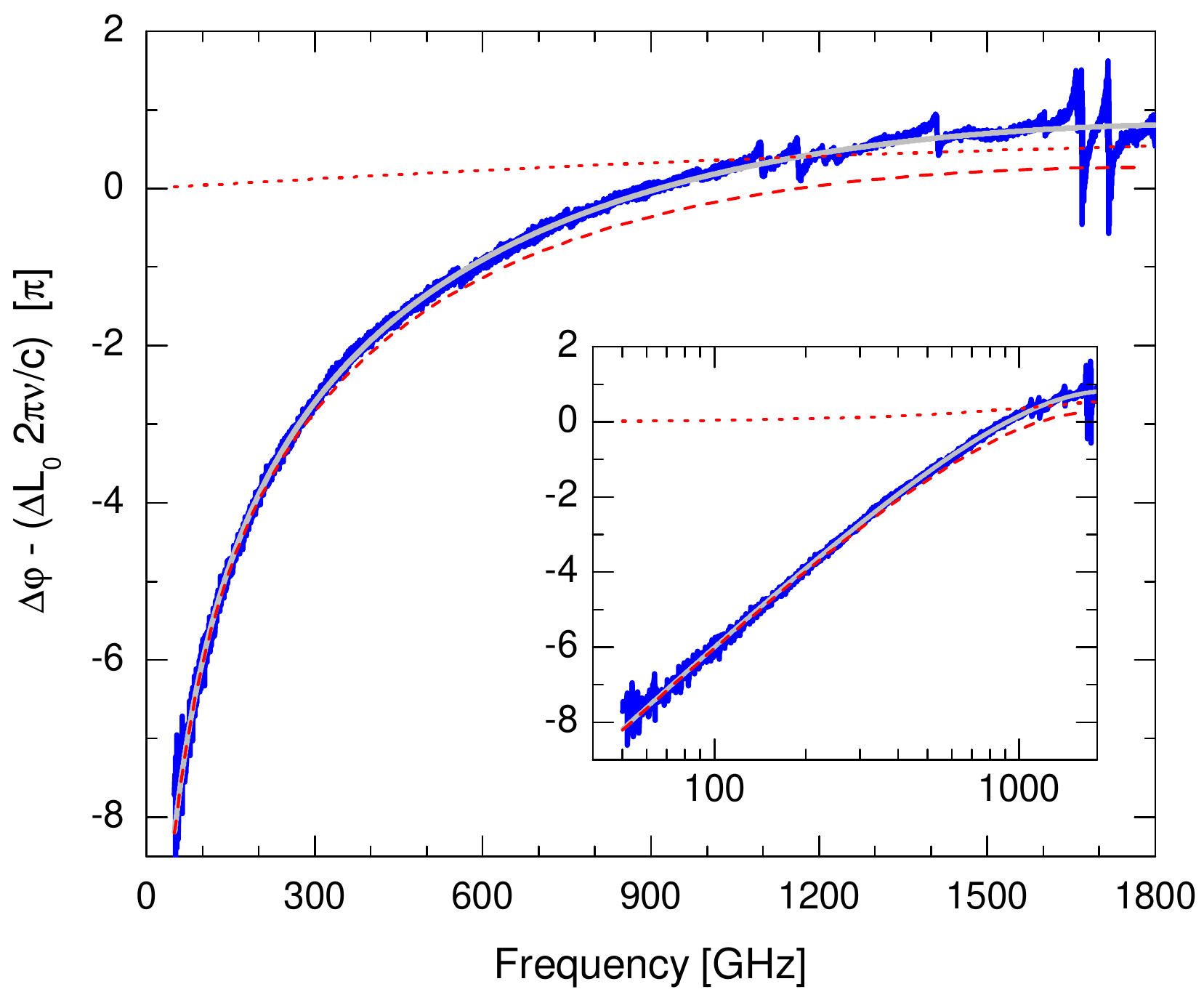}
\caption{Comparison of measured data and model, focussing on the non-linear contributions.
Blue: $\Delta \varphi(\nu)-\Delta L_0 \cdot 2\pi \nu /c$
for the data set with $\Delta L \! \approx \! -0.06$\,mm, cf.\ Fig.\ \ref{fig:phase}.
Dashed and dotted red lines depict $\Delta \varphi_{\rm an}$ and $\Delta \varphi_{RC}$, respectively,
which add up to $\Delta \varphi_{\rm mod}(\nu) - \Delta L_0 \cdot 2\pi \nu /c$ (grey),
cf.\ Eq.\ \ref{eq:model}.
Inset: same data on a logarithmic frequency scale.}
\label{fig:phaseminDL}
\end{figure}

\textit{(3) Terahertz path:}
The terahertz path length $L_{\rm THz}$ and the effective fiber lengths $L_{\rm Tx}$ and $L_{\rm Rx}$
constitute the optical path-length difference
$\Delta L$, contributing a term $\Delta L \cdot \frac{2\pi\nu}{c}$ to $\Delta \varphi(\nu)$,
see Eqs.\ \ref{eq:DL} and \ref{eq:beatphase_2laser}.
Due to water vapor in the terahertz path and standing waves between, e.g., the photomixers,
$L_{\rm THz}$ effectively depends on the frequency $\nu$.
For the discussion of the overall behavior of $\Delta \varphi(\nu)$, these effects are small compared
to the contribution $\Delta \varphi_{\rm an}(\nu)$ of the antenna.
Therefore, we first consider a constant value of $L_{\rm THz}$ and come back to these smaller effects below.

Having addressed all the different contributions, we derive
a simple model for the overall behavior of $\Delta \varphi(\nu)$ by taking into account
the antenna contribution (cf.\ Eq.\ \ref{eq:antenna}),
the photomixer impedance (cf.\ Eq.\ \ref{eq:atanphi}), and
a constant optical path-length difference $\Delta L_0$,
\begin{equation}
\Delta \varphi_{\rm mod}(\nu)  =
\Delta \varphi_{\rm an}(\nu) + \Delta \varphi_{RC}(\nu) + \Delta L_0 \cdot \frac{2\pi\nu}{c} \, .
\label{eq:model}
\end{equation}
For a quantitative comparison with the experimental results, we use the given values of
the lifetime $\tau_{RC}$\,=\,$R_{\rm A} C$\,=\,0.1\,ps, the spiral growth rate $a$\,=\,0.2,
and $\nu_{\rm max}$\,=\,1.8\,THz.
For the radius $r_{\rm ar}$ of the active region, we employ $2\pi r_{\rm ar}$\,=\,$\lambda$
(cf.\ Sec.\ \ref{sec:ar}), leaving only two free parameters,
$\Delta L_0$ and a constant offset denoted by $\Delta \varphi_{\rm an}(\nu_{\rm max})$,
cf.\ Eq.\ \ref{eq:antenna}.
Surprisingly, this simple model is in excellent agreement with our experimental data of $\Delta \varphi(\nu)$,
see Fig.\ \ref{fig:phase}.
If we view $\gamma$\,=\,$\lambda/2\pi r_{\rm ar}$ as an additional fit parameter,
we find $\gamma$\,=\,0.997 and $\Delta \varphi_{\rm an}(\nu_{\rm max})$\,=\,0.26$\pi$.

In order to highlight the non-linear terms $\Delta \varphi_{\rm an}(\nu)$ and $\Delta \varphi_{RC}(\nu)$,
we compare $\Delta \varphi(\nu) - \Delta L_0 \cdot 2\pi \nu /c$
with $\Delta \varphi_{\rm an}(\nu) + \Delta \varphi_{RC}(\nu)$
in Fig.\ \ref{fig:phaseminDL}.
The antenna contribution $\Delta \varphi_{\rm an}(\nu)$ clearly dominates since
it changes by roughly $6\pi$ between 100\,GHz and 1\,THz.

\begin{figure}[tb]
\centering
\includegraphics[width=.8\columnwidth]{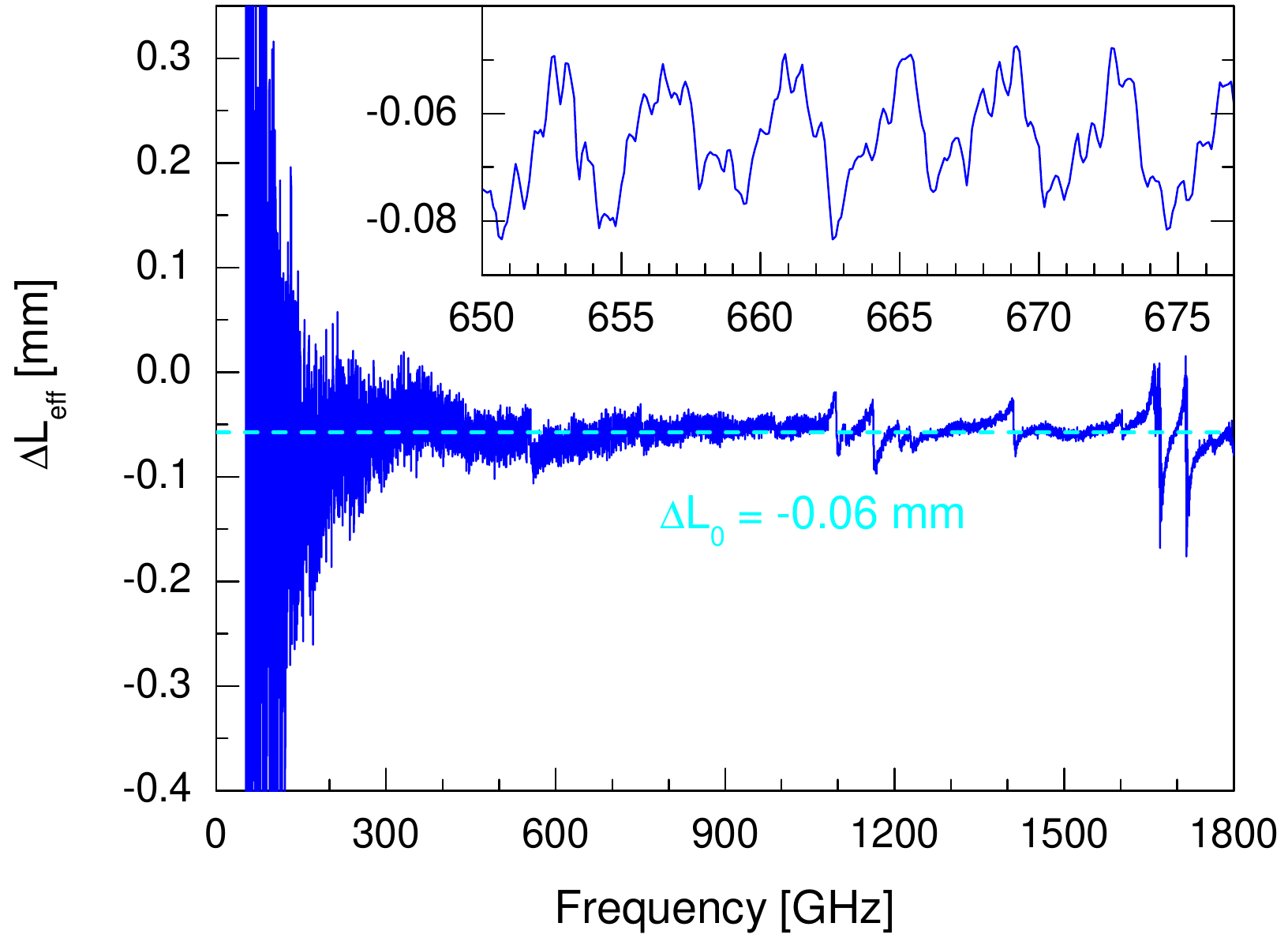}
\caption{Effective optical path-length difference $\Delta L_{\rm eff}(\nu)$ for the data set with $\Delta L \! \approx \! -0.06$\,mm,
cf.\ Eq.\ \ref{eq:DLeff} and Fig.\ \ref{fig:phase}.
Inset: same data on a larger scale, showing the 4.07\,GHz modulation
stemming from the Si lenses of the photomixers
as well as the 0.7\,GHz modulation caused by standing waves
between the photomixers for $L_{\rm THz} \! \approx \! 22$\,cm.}
\label{fig:DL}
\end{figure}

Finally, we define an effective, frequency-dependent optical path-length difference
$\Delta L_{\rm eff}(\nu)$ and a corrected phase difference $\Delta \varphi_{\rm corr}(\nu)$
by subtracting the two dominant non-linear terms from $\Delta \varphi(\nu)$,
\begin{eqnarray}
 \Delta L_{\rm eff}(\nu) \cdot \frac{2\pi \nu}{c} &=& \Delta \varphi_{\rm corr}(\nu) \\
\nonumber   &=& \Delta \varphi(\nu) - \Delta \varphi_{\rm an}(\nu) - \Delta \varphi_{RC}(\nu) \, .
\label{eq:DLeff}
\end{eqnarray}
The result is shown in Fig.\ \ref{fig:DL}.
The average of the effective optical path-length difference $\Delta L_{\rm eff}(\nu)$ equals $\Delta L_0$,
while the frequency dependence of $\Delta L_{\rm eff}(\nu)$ and accordingly of $\Delta \varphi_{\rm corr}(\nu)$
contains all deviations between the measured $\Delta \varphi(\nu)$ and $\Delta \varphi_{\rm mod}(\nu)$.
Due to the excellent agreement between $\Delta \varphi(\nu)$ and $\Delta \varphi_{\rm mod}(\nu)$,
the frequency dependence of $\Delta L_{\rm eff}(\nu)$ highlights the smaller contributions that we have
neglected thus far, i.e., the effective frequency dependence of $L_{\rm THz}$.
We identify three main contributions:
(a) Standing waves within the Si lenses give rise to a modulation of $\Delta L_{\rm eff}$
with a period of about 4.1\,GHz, see inset of Fig.\ \ref{fig:DL}.
(b) Standing waves between the two photomixers separated by $L_{\rm THz} \! \approx \! 22$\,cm
cause a modulation period of 0.7\,GHz\,$\approx$\,$c/$0.4\,m.
These periodic features are well resolved in the Fourier-transformed data, see Sec.\ \ref{sec:FT}.
(c) We observe resonant absorption features of water vapor.\cite{Pickett98,Grischkowsky2013}
The absorption lines are very well resolved even for this comparably short path in air.
Very roughly, the absorption line at 557\,GHz is expected to cause a peak-to-peak change
of about $2 \cdot 10^{-4}$ of the refractive index of air.\cite{Grischkowsky2013}
For $L_{\rm THz} \! \approx \!22$\,cm, this corresponds to about 40\,$\mu$m peak-to-peak,
in rough agreement with our data.

\section{Active region}
\label{sec:ar}

The antenna radiates most strongly from a region in which the two spiral arms radiate in phase,
giving rise to constructive interference.\cite{Kaiser60,McFaddenPhD}
For the radius $r_{\rm ar}$ of the active region, we consider a spot with $r_{\rm ar}$\,=\,$(r_{+} + r_{-})/2$
which is located between the two neighboring arms with radii $r_-$ and $r_{+}$\,=\,$r_- \cdot e^{a\pi}$, respectively.
There, the path length of the two neighboring arms differs by $l(r_{+})-l(r_{-})$\,=\,$l(\alpha+\pi)-l(\alpha)$,
cf.\ Eq.\ \ref{eq:lengthspiral}.
The antenna is fed in a balanced way, thus the currents in the two arms are out-of-phase at $\pm r_{\rm min}$.
Constructive interference occurs if the path-length difference between the two arms equals $\lambda/2$,
compensating for the initial phase shift of $\pi$. With Eqs.\ \ref{eq:spiral} and \ref{eq:lengthspiral} we find
\begin{equation}
\frac{\lambda}{2} = l(\alpha+\pi) - l(\alpha) = \frac{\sqrt{1+a^2}}{a} (r_{-} - r_{\rm min}) \cdot (e^{a\pi}-1) \, .
\label{eq:ar_pathdiff}
\end{equation}
Neglecting $r_{\rm min} \! \approx \! 10$\,$\mu$m\,$\ll \lambda$, the radius of the active region amounts to
\begin{equation}
r_{\rm ar} = \frac{r_{+} + r_{-}}{2} = \frac{a}{4\sqrt{1+a^2}} \frac{e^{a\pi}+1}{e^{a\pi}-1} \cdot \lambda \, .
\label{eq:ar}
\end{equation}
For a vanishing growth rate $a\rightarrow 0$, this yields $2\pi r_{\rm ar}$\,=\,$\lambda$.
For $a$\,=\,0.2, we find $2 \pi r_{\rm ar} \! \approx \! 1.01 \lambda$, in excellent agreement with our experimental result.

A theoretical study of the
radiated power density as a function of $2\pi r/\lambda$ has been reported in Ref.\ [\onlinecite{Piksa12}]
for planar log-spiral antennae with different growth rates.
For $a\! \approx \! 0.18 \! \approx 1/\tan(80^\circ)$, the power density shows a rather broad, asymmetric peak at
about $2\pi r$\,=\,$\lambda/2$ and decreases only slowly towards higher values of $r$.
Integrating the contributions from $2\pi r$\,=\,$0.2\lambda$ to $2\lambda$ yields a first moment of about
$0.9\lambda$, in fair agreement with our results.

\section{Quasi-time-domain analysis}
\label{sec:FT}

\begin{figure}[tb!]
\centering
\includegraphics[width=1.0\columnwidth]{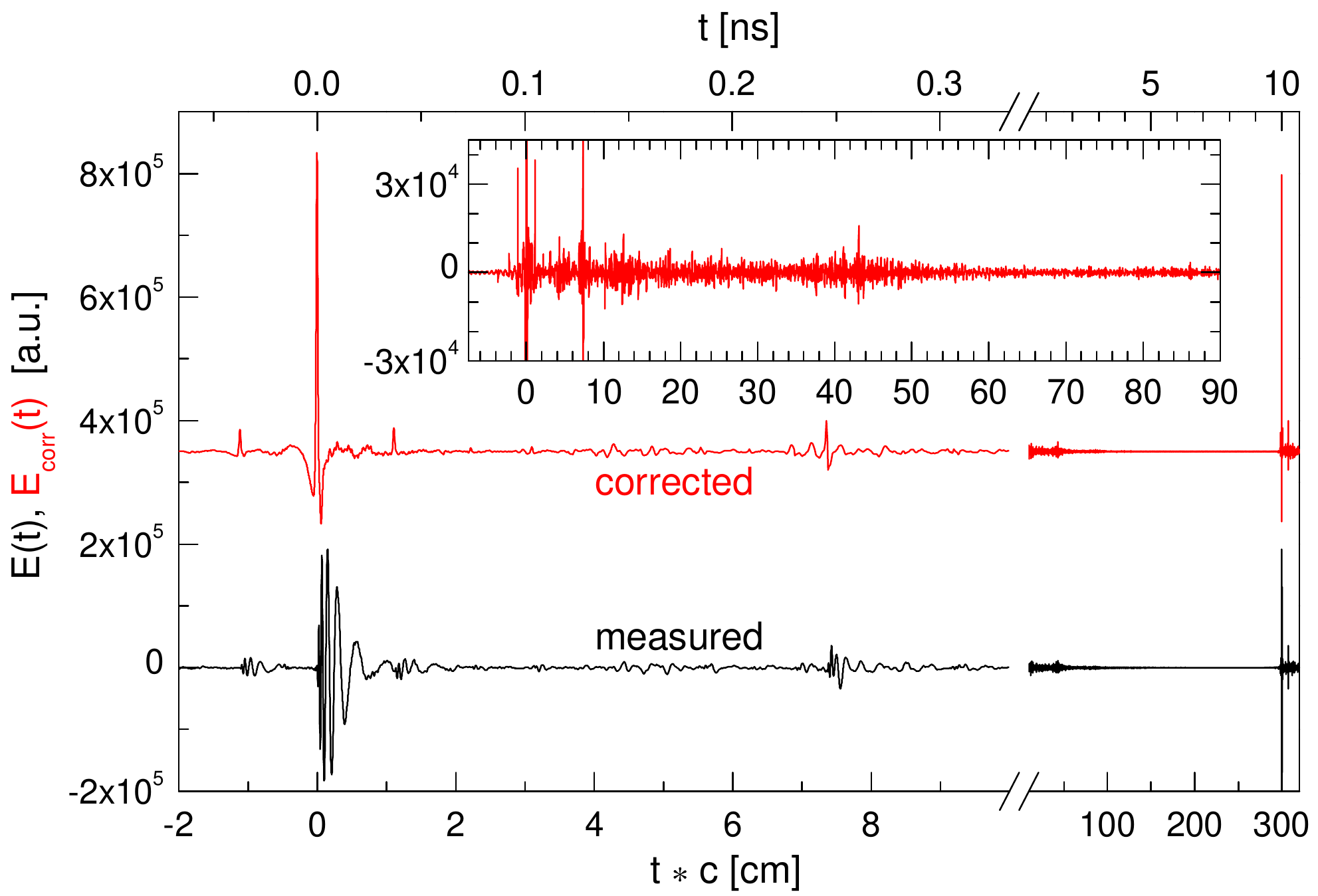}
\caption{Fourier series $E(t)$ of the data (black) measured up to 1.8\,THz for $\Delta L \! \approx \! -0.06$\,mm
(cf.\ Fig.\ \ref{fig:phase}) and after correction (red), cf.\ Eq.\ \ref{eq:phicorr}.
Data are offset for clarity.
A frequency step width of 100\,MHz yields $E(t)$ with a period $T \! \approx \! 300$\,cm/$c$.
Note the change of scale at 10\,cm.
Inset: Corrected data on a different scale.
The feature which is shifted by 7.36\,cm with respect to the main peak results from
standing waves in the Si lenses, the peak at 43.1\,cm corresponds to $2\cdot L_{\rm THz}$
and is caused by standing waves between the two photomixers.}
\label{fig:FT}
\end{figure}

Knowing amplitude $E_{\rm THz}(\nu)$ and phase difference $\Delta \varphi(\nu)$ for the discrete set of
frequency points of a certain measurement, one can easily calculate the Fourier series as a function of the time $t$,
\begin{equation}
 E(t) = \sum_{\nu} E_{\rm THz}(\nu) \, \cos\left(2\pi \nu t - \Delta \varphi(\nu)\right) \, ,
\end{equation}
an approach which has been called quasi-time-domain analysis.\cite{Scheller2009}
The Fourier series is equivalent to an interferogram or to the waveform in a time-domain
terahertz experiment, where all frequencies are measured simultaneously.

In the Fourier series, the main peak is expected at $t$\,=\,$\Delta L/c$.
The measured data of $E(t)$ (black line in Fig.\ \ref{fig:FT}) do not show
well-defined peak positions, but are strongly asymmetric around any peak.
This is the typical shape of a down-chirp signal, in which the higher frequencies arrive first,
reflecting the strongly frequency-dependent group delay introduced by the antenna.
However, using the corrected phase difference
$\Delta\varphi_{\rm corr}(\nu)$ (cf.\ Eq.\ \ref{eq:DLeff}) in the Fourier series
\begin{equation}
 E_{\rm corr}(t) = \sum_{\nu} E_{\rm THz}(\nu) \, \cos\left(2\pi \nu t - \Delta \varphi_{\rm corr}(\nu)\right)
\label{eq:phicorr}
\end{equation}
removes the strong down-chirp and yields a pronounced ``pulse'' at the expected position (red line in Fig.\ \ref{fig:FT}).
The remaining peak width reflects the finite bandwidth of the experiment and in particular the strong decrease
of the amplitude with increasing frequency.
The corrected data also show a clear feature shifted by 7.36\,cm with respect to the main peak,
which is equivalent to a period of 4.07\,GHz. This feature reflects the periodic modulations shown in the inset
of Fig.\ \ref{fig:DL}, i.e., standing waves in the Si lenses.
Peaks at $t\cdot c$\,=\,$\pm 1.1$\,cm are due to standing waves in the tapered amplifier.
These occur before the optical path is split into two arms and thus do not contribute to $\Delta \varphi$,
i.e., they are only observed in the amplitude.
The feature at 43.1\,cm reflects standing waves between the two photomixers with $L_{\rm THz}\! \approx \! 22$\,cm.
We also observe an overtone at about 86\,cm.
Using the corrected data clearly facilitates
the detection and precise determination of such features in the Fourier series.

\section{Group delay and uncertainty of the phase}
\label{sec:gr}

\begin{figure}[tb!]
\centering
\includegraphics[width=.85\columnwidth]{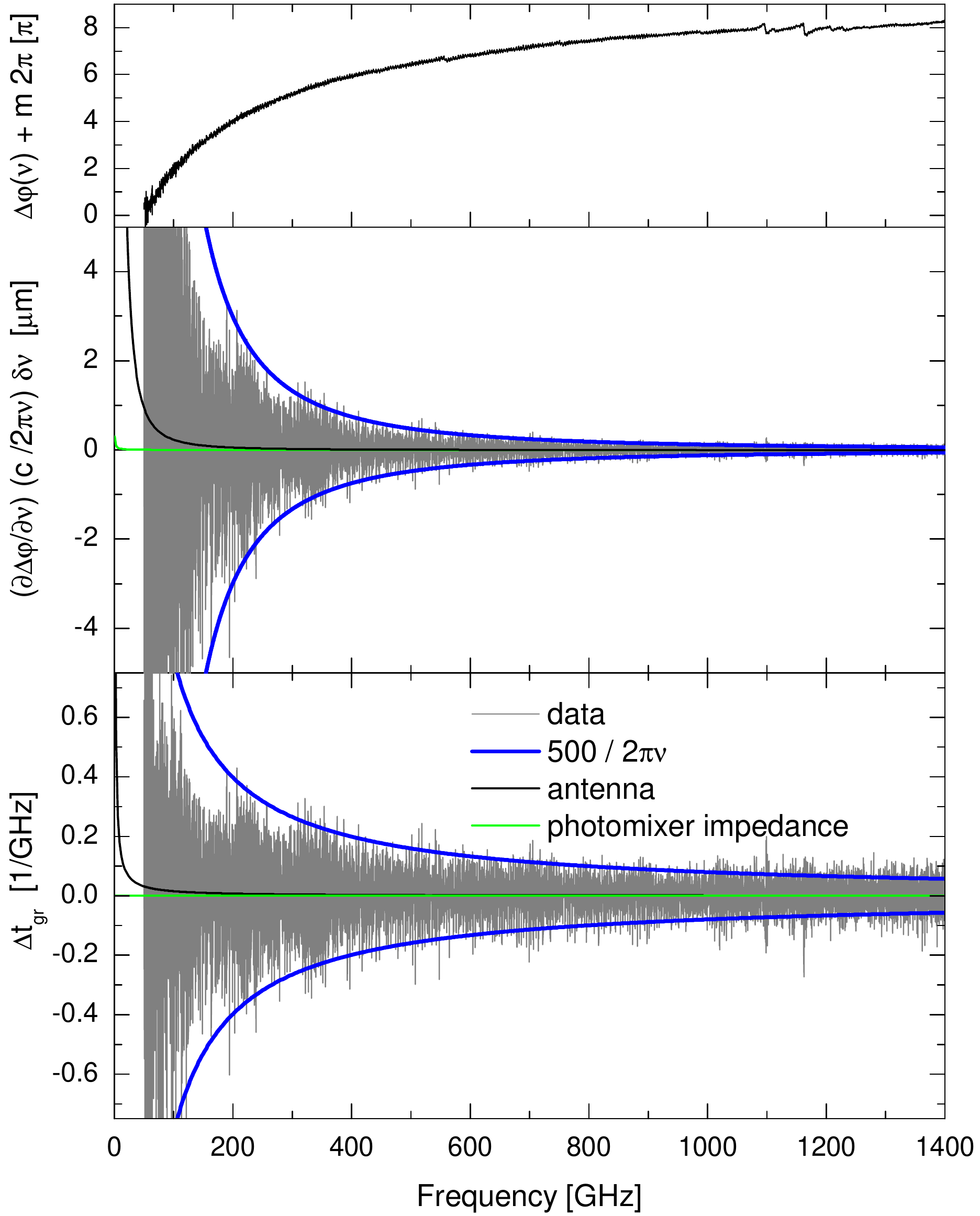}
\caption{Top: phase difference $\Delta \varphi(\nu)$ for $\Delta L \! \approx \! -0.06$\,mm, cf.\ Fig.\ \ref{fig:phase},
measured with a frequency step width of 100\,MHz.
Bottom: group delay difference $\Delta t_{\rm gr}$\,=\,$(\partial \Delta \varphi/\partial \nu)/2\pi$ compared to the
contributions of the antennae, $(\partial \Delta \varphi_{\rm an}/\partial \nu)/2\pi$ (black), and of the photomixer
impedance, $(\partial \Delta \varphi_{RC}/\partial \nu)/2\pi$ (green).
Blue: rough estimate of the envelope.
Middle: Same data as in the bottom panel, multiplied by $(c/\nu) \delta \nu$ with $\delta \nu$\,=\,5\,MHz,
for comparison with a length change $\delta L$,
cf.\ Eq.\ \ref{eq:delnu}.}
\label{fig:phase_deriv_700}
\end{figure}

The uncertainty $\delta \varphi$ of the measured phase difference $\Delta \varphi(\nu)$ depends on
the experimental uncertainties of $\Delta L$ and $\nu$,
\begin{equation}
 \delta \varphi =  \frac{\partial \Delta \varphi}{\partial \nu} \cdot \delta \nu
 + \frac{2\pi\nu}{c} \cdot \delta L  \, .
\label{eq:phase_stab}
\end{equation}
In our setup,
the optical path-length difference is typically stable to within $\delta L$\,=\,$\pm 5$\,$\mu$m.\cite{Roggenbuck12}
The line width of the beat signal of the two widely tunable lasers amounts to about 5\,MHz,
while a long-term frequency stability of better than 20\,MHz over 24\,h was observed.\cite{Deninger08}
On the time scale of less than 1\,h, the frequency stability is better than $\delta \nu$\,=\,5\,MHz.
The quantitative understanding of $\Delta \varphi(\nu)$ achieved in the previous sections allows us to discuss
the importance of the different contributions to the group delay difference
$\Delta t_{\rm gr} \propto \partial \Delta \varphi/\partial \nu$, i.e., to the first term on the right hand side.
We consider (cf.\ Eq.\ \ref{eq:DLeff})
\begin{equation}
 \frac{\partial \Delta \varphi}{\partial \nu} =
 \frac{\partial \Delta \varphi_{\rm an}}{\partial \nu} +
 \frac{\partial \Delta \varphi_{RC}}{\partial \nu} +
 \Delta L_{\rm eff} \cdot \frac{2\pi}{c} +
 \frac{\partial \Delta L_{\rm eff}}{\partial \nu}\cdot \frac{2\pi\nu}{c} \,\,\, .
\label{eq:dphidnu}
\end{equation}
For the first two terms we find
\begin{eqnarray}
\nonumber
\frac{\partial \Delta \varphi_{\rm an}}{\partial \nu} &=& \frac{2 \sqrt{1+a^2}}{a}\cdot \left( \frac{1}{\nu} - \frac{1}{\nu_{\rm max}} \right) \\
\frac{\partial \Delta \varphi_{RC}}{\partial \nu} &=& \frac{4 \pi \tau_{RC}}{1 +(2\pi \nu \tau_{RC})^2} \, .
\label{eq:deriv_an}
\end{eqnarray}
The term $\Delta L_{\rm eff} \cdot 2\pi/c$ dominates for large values of $\Delta L$ but can be suppressed by choosing a small $\Delta L$.
For instance, it amounts to about 0.2/GHz for $\Delta L$\,=\,1\,cm.
The data shown in Fig.\ \ref{fig:phase_deriv_700} was measured with $\Delta L \! \approx \! -0.006$\,cm (cf.\ Fig.\ \ref{fig:DL}),
thus the third term in Eq.\ \ref{eq:dphidnu} can be neglected.
The comparison with experimental data in Fig.\ \ref{fig:phase_deriv_700} shows that
the first two terms can equally be neglected.
Although the contributions of $\Delta \varphi_{\rm an}$ and $\Delta \varphi_{RC}$ dominate the overall behavior
of $\Delta \varphi(\nu)$, they are both negligible with respect to
the derivative $\partial \Delta \varphi/\partial \nu$.
This derivative is dominated by the remaining term $\propto \partial \Delta L_{\rm eff}/\partial \nu$,
i.e., by the contribution of standing waves within the Si lenses and between the two photomixers.
Empirically, we find that the envelope of $\partial \Delta \varphi/\partial \nu$ is roughly
described by $f(\nu)$\,=\,$\pm 500/\nu$ for $\Delta L \! \approx \! -0.06$\,mm and $L_{\rm THz} \! \approx \! 22$\,cm,
see bottom panel of Fig.~\ref{fig:phase_deriv_700}.

In order to compare the effects of frequency uncertainty versus length drift, we consider
\begin{equation}
 \delta \varphi \cdot \frac{c}{2\pi\nu} =
 \frac{\partial \Delta \varphi}{\partial \nu} \cdot \frac{c}{2\pi\nu} \cdot \delta \nu
 +  \delta L  \, .
 \label{eq:delnu}
\end{equation}
With $\delta \nu$\,=\,5\,MHz and the experimental result for the envelope of $\pm 500/\nu$,
the first term roughly yields
0.12\,$\mu$m $\cdot$ (THz/$\nu)^2$, which amounts to
3\,$\mu$m at 200\,GHz or
0.75\,$\mu$m at 400\,GHz,
see middle panel of Fig.\ \ref{fig:phase_deriv_700}.
The typical length drift observed in our setup equals $\pm$\,5\,$\mu$m.\cite{Roggenbuck12}
We conclude that
for $\delta L$\,=\,$\pm$5\,$\mu$m, $\Delta L \lesssim 1$\,cm, and frequencies above about 200\,GHz,
the uncertainty $\delta \varphi/\nu$ mainly depends on the drift of the optical path-length difference,
in agreement with the experimental results discussed in Ref.\ [\onlinecite{Roggenbuck12}].

\section{Conclusions}

We investigated the phase difference $\Delta \varphi(\nu)$
between transmitter arm and receiver arm in cw terahertz spectroscopy based on
photomixers with ultra-wideband log-spiral antennae.
We find that $\Delta \varphi(\nu)$ and the group delay difference
$\Delta t_{\rm gr} \propto \partial \Delta \varphi/\partial \nu$ are dominated by different terms.
The overall behavior of $\Delta \varphi(\nu)$ is quantitatively described by taking into account
three different contributions. The optical path-length difference gives rise to a term
linear in frequency, while the radiation characteristics of the log-spiral antennae
and the photomixer impedance cause deviations from this linear behavior.
The contribution $\Delta \varphi_{\rm an}(\nu)$
of the log-spiral antennae is very well described by a simple model which assumes
that the antennae effectively radiate and receive in an active region in which the circumference equals $\lambda$.
Correcting for the group delay of the antennae and photomixers strongly facilitates an analysis
of the Fourier-transformed spectra.
In contrast to $\Delta \varphi(\nu)$, the derivative $\partial \Delta \varphi/\partial \nu$ is dominated by the
contribution of standing waves, i.e., periodic modulations of $\Delta \varphi(\nu)$.
In combination with a finite frequency error, these standing waves may affect the experimental uncertainty
$\delta \varphi$, but typically their contribution can be neglected in comparison to the effect of a
drift of the optical path-length difference.
Nevertheless it is advisable to suppress standing waves with a small modulation period, in particular
for measurements at low frequencies.

\end{document}